\input{psfig}
\documentstyle[12pt,psfig]{article}
\catcode`\@=11

\@addtoreset{equation}{section}
\def\theequation{\arabic{equation}}
\def\theequation{\thesection\arabic{equation}}

\def\NPB#1#2#3{{\it Nucl.~Phys.} {\bf{B#1}} (19#2) #3}
\def\PLB#1#2#3{{\it Phys.~Lett.} {\bf{B#1}} (19#2) #3}
\def\PRD#1#2#3{{\it Phys.~Rev.} {\bf{D#1}} (19#2) #3}
\def\PRL#1#2#3{{\it Phys.~Rev.~Lett.} {\bf{#1}} (19#2) #3}

\def\@normalsize{\@setsize\normalsize{15pt}\xiipt\@xiipt
\abovedisplayskip 14pt plus3pt minus3pt%
\belowdisplayskip \abovedisplayskip
\abovedisplayshortskip  \z@ plus3pt%
\belowdisplayshortskip  7pt plus3.5pt minus0pt}
\def\small{\@setsize\small{13.6pt}\xipt\@xipt
\abovedisplayskip 13pt plus3pt minus3pt%
\belowdisplayskip \abovedisplayskip
\abovedisplayshortskip  \z@ plus3pt%
\belowdisplayshortskip  7pt plus3.5pt minus0pt
\def\@listi{\parsep 4.5pt plus 2pt minus 1pt
            \itemsep \parsep
            \topsep 9pt plus 3pt minus 3pt}}

\def\underline#1{\relax\ifmmode\@@underline#1\else
        $\@@underline{\hbox{#1}}$\relax\fi}
\@twosidetrue
\relax

\catcode`@=12

\catcode`\@=11

\def\section{\@startsection{section}{1}{\z@}{3.5ex plus 1ex minus
   .2ex}{2.3ex plus .2ex}{\large\bf}}
\def\thesection{\arabic{section}.}

\def\ps@headings{\def\@oddfoot{}\def\@evenfoot{}
\def\@oddhead{\hbox{}\hfill
        \makebox[.5\textwidth]{\raggedright\ignorespaces --\thepage{}--
        \hfill }}
\def\@evenhead{\@oddhead}
\def\subsectionmark##1{\markboth{##1}{}}
}

\ps@headings

\catcode`\@=12

\relax

\def\figcap{\section*{Figure Captions\markboth
        {FIGURECAPTIONS}{FIGURECAPTIONS}}\list
        {Fig. \arabic{enumi}:\hfill}{\settowidth\labelwidth{Fig. 999:}
        \leftmargin\labelwidth
        \advance\leftmargin\labelsep\usecounter{enumi}}}
 \relax
\def\tablecap{\section*{Table Captions\markboth
        {TABLECAPTIONS}{TABLECAPTIONS}}\list
        {Table \arabic{enumi}:\hfill}{\settowidth\labelwidth{Table 999:}
        \leftmargin\labelwidth
        \advance\leftmargin\labelsep\usecounter{enumi}}}
 \relax
\def\reflist{\section*{References\markboth
        {REFLIST}{REFLIST}}\list
        {[\arabic{enumi}]\hfill}{\settowidth\labelwidth{[999]}
        \leftmargin\labelwidth
        \advance\leftmargin\labelsep\usecounter{enumi}}}
 \relax

\catcode`\@=11

\def\marginnote#1{}
\newcount\hour
\newcount\minute
\newtoks\amorpm
\hour=\time\divide\hour by60
\minute=\time{\multiply\hour by60 \global\advance\minute by-
\hour}
\edef\standardtime{{\ifnum\hour<12 \global\amorpm={am}%
    \else\global\amorpm={pm}\advance\hour by-12 \fi
    \ifnum\hour=0 \hour=12 \fi
    \number\hour:\ifnum\minute<100\fi\number\minute\the\amorpm}}
\edef\militarytime{\number\hour:\ifnum\minute<100\fi\number\minute}
\def\draftlabel#1{{\@bsphack\if@filesw {\let\thepage\relax
  \xdef\@gtempa{\write\@auxout{\string
    \newlabel{#1}{{\@currentlabel}{\thepage}}}}}\@gtempa
    \if@nobreak \ifvmode\nobreak\fi\fi\fi\@esphack}
     \gdef\@eqnlabel{#1}}
\def\@eqnlabel{}
\def\@vacuum{}
\def\draftmarginnote#1{\marginpar{\raggedright\scriptsize\tt#1}}
\def\draft{\oddsidemargin -.5truein
        \def\@oddfoot{\sl preliminary draft \hfil
        \rm\thepage\hfil\sl\today\quad\militarytime}
        \let\@evenfoot\@oddfoot \overfullrule 3pt
        \let\label=\draftlabel
        \let\marginnote=\draftmarginnote
   
\def\@eqnnum{(\theequation)\rlap{\kern\marginparsep\tt\@eqnlabel}%
\global\let\@eqnlabel\@vacuum}  }
\def\preprint{\twocolumn\sloppy\flushbottom\parindent 1em
        \leftmargini 2em\leftmarginv .5em\leftmarginvi .5em
        \oddsidemargin -.5in    \evensidemargin -.5in
        \columnsep 15mm \footheight 0pt
        \textwidth 250mmin      \topmargin  -.4in
        \headheight 12pt \topskip .4in
        \textheight 175mm
        \footskip 0pt
        
\def\@oddhead{\thepage\hfil\addtocounter{page}{1}\thepage}
        \let\@evenhead\@oddhead \def\@oddfoot{} \def\@evenfoot{} 
}
\def\titlepage{\@restonecolfalse\if@twocolumn\@restonecoltrue\onecolumn
     \else \newpage \fi \thispagestyle{empty}\c@page\z@
        \def\thefootnote{\fnsymbol{footnote}} }
\def\endtitlepage{\if@restonecol\twocolumn \else  \fi
        \def\thefootnote{\arabic{footnote}}
        \setcounter{footnote}{0}}  
\catcode`@=12
\relax

\def\ps@headings{\def\@oddfoot{}\def\@evenfoot{}
\def\@oddhead{\hbox{}\hfill
        \makebox[.5\textwidth]{\raggedright\ignorespaces --\thepage{}--
        \hfill }}
\def\@evenhead{\@oddhead}
\def\subsectionmark##1{\markboth{##1}{}}
}

\ps@headings

\relax

\def\firstpage#1#2#3#4#5#6{
\begin{document}
\begin{titlepage}
\nopagebreak
\title{\begin{flushright}
        \vspace*{-1.8in}
        {\normalsize CERN-TH/97-260}\\[-9mm]
        {\normalsize CPTH-S562.0997}\\[-9mm]
        {\normalsize hep-ph/9710204}\\[4mm]
\end{flushright}
\vfill
{#3}}
\author{\large #4 \\[1.0cm] #5}
\maketitle
\vskip -7mm     
\nopagebreak 
\begin{abstract}
{\noindent #6}
\end{abstract}
\vfill
\begin{flushleft}
\rule{16.1cm}{0.2mm}\\[-3mm]
$^{\star}${\small Research supported in part by the EEC under TMR contract 
ERBFMRX-CT96-0090.}\\[-3mm] 
$^{\dagger}${\small Laboratoire Propre du CNRS UPR A.0014.}\\
CERN-TH/97-260\\
September 1997
\end{flushleft}
\thispagestyle{empty}
\end{titlepage}}

\def\simlt{\stackrel{<}{{}_\sim}}
\def\simgt{\stackrel{>}{{}_\sim}}
\newcommand{\dal}{\raisebox{0.085cm}
{\fbox{\rule{0cm}{0.07cm}\,}}}
\newcommand{\dt}{\partial_{\langle T\rangle}}
\newcommand{\dtbar}{\partial_{\langle\overline{T}\rangle}}
\newcommand{\al}{\alpha^{\prime}}
\newcommand{\mst}{M_{\scriptscriptstyle \!S}}
\newcommand{\mpl}{M_{\scriptscriptstyle \!P}}
\newcommand{\dv}{\int{\rm d}^4x\sqrt{g}}
\newcommand{\lv}{\left\langle}
\newcommand{\rv}{\right\rangle}
\newcommand{\ph}{\varphi}
\newcommand{\abar}{\overline{a}}
\newcommand{\sbar}{\,\overline{\! S}}
\newcommand{\xbar}{\,\overline{\! X}}
\newcommand{\fbar}{\,\overline{\! F}}
\newcommand{\zbar}{\overline{z}}
\newcommand{\dbar}{\,\overline{\!\partial}}
\newcommand{\tbar}{\overline{T}}
\newcommand{\taubar}{\overline{\tau}}
\newcommand{\ubar}{\overline{U}}
\newcommand{\ybar}{\overline{Y}}
\newcommand{\phb}{\overline{\varphi}}
\newcommand{\cm}{Commun.\ Math.\ Phys.~}
\newcommand{\prl}{Phys.\ Rev.\ Lett.~}
\newcommand{\pr}{Phys.\ Rev.\ D~}
\newcommand{\pl}{Phys.\ Lett.\ B~}
\newcommand{\ibar}{\overline{\imath}}
\newcommand{\jbar}{\overline{\jmath}}
\newcommand{\np}{Nucl.\ Phys.\ B~}
\newcommand{\F}{{\cal F}}
\renewcommand{\L}{{\cal L}}
\newcommand{\A}{{\cal A}}
\newcommand{\e}{{\rm e}}
\newcommand{\be}{\begin{equation}}
\newcommand{\en}{\end{equation}}
\newcommand{\ba}{\begin{eqnarray}}
\newcommand{\dslash}{{\not\!\partial}}
\newcommand{\ea}{\end{eqnarray}}
\newcommand{\gsi}{\,\raisebox{-0.13cm}{$\stackrel{\textstyle
>}{\textstyle\sim}$}\,}
\newcommand{\lsi}{\,\raisebox{-0.13cm}{$\stackrel{\textstyle
<}{\textstyle\sim}$}\,}
\date{}
\firstpage{3118}{IC/95/34}
{\large\bf Millimetre-Range Forces in Superstring Theories with Weak-Scale  
Compactification} 
{I. Antoniadis$^{\,a,b}$, S. Dimopoulos$^{\,a,c}$ and G. Dvali$^{\,a}$}
{\normalsize\sl
$^a$TH-Division, CERN, CH-1211 Geneva 23, Switzerland\\[-3mm]
\normalsize\sl$^b$ Centre de Physique Th{\'e}orique, 
Ecole Polytechnique,$^\dagger$ {}F-91128 Palaiseau, France\\[-3mm]
\normalsize\sl
$^c$ Physics Department, Stanford University, Stanford, CA 94305, USA}
{We show that theories in which supersymmetry is broken via  
Scherk-Schwarz compactification at the weak scale, possess  
at least one scalar particle with Compton wavelength in the  
millimetre range, which mediates a force with strength 1/3 of gravity.
Such forces are going to be explored in upcoming experiments using
micro-electromechanical systems or cantilever technology. We also 
present a simple way of understanding some decoupling aspects of these 
theories by analogy with finite-temperature field theory.}


\section{Light scalars and theories of supersymmetry breaking}

One of the most outstanding challenges of supersymmetric theories is  
the problem of the breaking of supersymmetry. It is directly connected  
with the cosmological constant problem for which no solution is in  
sight. In spite of this big hole in our theoretical understanding, the  
idea of softly broken supersymmetric theories
\cite{dg} has allowed us to bypass these questions  
and to study phenomenological consequences of supersymmetric theories  
in the past 16 years.
In the mean time, three distinct classes of theories have emerged for  
the breaking of supersymmetry:\\
- Gravity-mediated theories.\\
- Gauge-mediated theories.\\
- Theories with weak-scale compactification.

A main difference between these approaches is the scale of  
supersymmetry breaking. In gravity-mediated theories it is $10^{13}$  
GeV. In gauge-mediated ones it is anywhere from over $10$ TeV to  
$10^{12}$ GeV. In theories of weak-scale compactification the  
supersymmetry breaking scale is around the weak scale, $\sim 1$ TeV \cite{a}.  
The different supersymmetry breaking scales imply important   
differences in the pattern of sparticle masses; these differences will  
be tested in the next decade at the LHC. The main purpose of this paper  
is to point out a possible difference between these theories that may  
emerge at very low energies, in the near future, at ``tabletop''  
experiments costing orders of magnitude less than the LHC. These involve  
macroscopic gravitational strength forces and arise as follows.

In supersymmetric or superstring theories one often encounters scalar  
particles, called moduli, which parametrize the size and the shape of the
extra compact dimensions. They couple with gravitational strength and are 
massless to all orders in perturbation theory. The moduli may get masses of 
the order of the Planck mass from non-perturbative effects; in this case 
they are not relevant to our considerations. It is, however, quite possible 
that they do not get masses until supersymmetry is broken. In this case 
their mass, because all their couplings are gravitationally suppressed, 
will be of order
\be
m_{\rm moduli}\, \sim\, {F\over M_p}\, ,
\label{mmod}
\en
where $M_p=2.4\times 10^{18}$ GeV is the reduced Planck mass, and
$F$ is the scale of supersymmetry breaking with dimensions of  
mass-squared. In gravity-mediated theories these moduli have weak  
scale masses and, since they are gravitationally coupled, they are not  
relevant to phenomenology. In gauge-mediated theories, if the scale  
of supersymmetry breaking is 10 TeV, the moduli can be quite light and  
have Compton wavelengths of  a few microns \cite{dgiu}.
Finally, in theories with weak-scale compactification, the moduli can 
have Compton wavelengths of the order 
of a millimetre. In fact, taking the scale of supersymmetry breaking from 
$F=(1\ {\rm TeV})^2$ to $(10\ {\rm TeV})^2$ we find a range of $\sim$ 1 mm 
to 10 $\mu$m for the range of the Compton wavelengths of the moduli \cite{fkz}.
Since the moduli are gravitationally coupled, they would lead to apparent 
deviations in Newton's universal law of gravitation at these distances.

Of course, measuring gravitational strength forces at such short  
distances is quite challenging. Nevertheless, a number of recent  
heroic experimental proposals aim at looking for precisely such forces  
in this range \cite{price,kapitulnik}. The most important background is  
the Van der Walls force. The Van der Walls
and gravitational forces between two atoms are equal to each other  
when the atoms are about 100 $\mu$m apart. Since the Van der Walls  
force falls off as the 7th power of the distance, it rapidly becomes  
negligible compared to gravity at distances exceeding 100 $\mu$m.
Therefore, theories of weak-scale compactification offer us the  
distinct possibility of measuring gravitational-strength forces at  
distances in the millimetre range. In fact, as we shall  
argue later, these theories are very likely to contain at least  
one modulus of such large Compton wavelength. This modulus is the one  
associated with the size of the ``large'' compact dimension of the  
order of the weak scale. It is essentially the logarithm of the radius  
of the large extra dimension. As we will show later this modulus  
couples with a strength that is 1/3 that of gravity. In addition it is  
quite possible that these theories possess other moduli which couple with  
strength up to $\sim$ 2000 times gravity \cite{tv}.

The outline of this paper is as follows. We first sketch, in Section 2, 
the theories of weak-scale compactification and subsequently show, in 
Section 3, how the ultraviolet behaviour of these theories is tame, 
by using the analogy with finite-temperature field theory. As a result, 
although the soft supersymmetry breaking masses are ``hard'', the vacuum 
energy density has no power law sensitivity to the cutoff $M_p$. Next 
and most important, in Section 4, we compute the mass range and the coupling 
of the large radius modulus to matter. Finally, in Section 5, we end with a 
summary of the special properties of weak-scale compactification theories.

Those results which are relevant for experiment are summarized in the  
figure. It depicts the strength of the modulus force relative to  
gravity versus distance. There we show all the presently  
allowed/excluded regions and the claimed capabilities of upcoming  
experiments. Particularly noteworthy are the horizontal lines  
corresponding to the ``large radius modulus'' and the dilaton. These  
theories can  have several light moduli; their couplings are expected  
to lie between those of the dilaton and the large radius modulus.

\section{Supersymmetry breaking with weak-scale compactification}

A large compact dimension at the TeV range is one of the few general low-energy 
predictions of {\it perturbative} superstring theory, which relates its size $R$
with the scale of supersymmetry breaking. In fact, superstring theories 
do not possess any small dimensionful parameters associated with supersymmetry
breaking \cite{susybr}. As a result, there are two ways to get a weak scale 
supersymmetry breaking mass in string theory. One is dimensional transmutation, 
which requires non-perturbative phenomena. The second is large radius 
compactification with $R\sim {\rm TeV}^{-1}$.

The main problem of such a large dimension is that all couplings become strong
very rapidly above the decompactification scale, because of the contribution 
of the infinite tower of Kaluza-Klein (KK) excitations, which change the 
logarithmic evolution with the energy into a power, as expected in an effective 
higher-dimensional non-renormalizable field theory. Fortunately, this problem,
which exists even before supersymmetry breaking is turned on,
can be avoided in a particular class of string models that include the simple
case of orbifold compactifications \cite{a}.

The spectrum of these theories consists of two sectors. The untwisted sector
contains the states of the toroidal compactification, which are invariant under
the action of the orbifold group. These are for instance all gauge multiplets.
At energies above the decompactification scale, but much below the string scale, 
they are characterized by an integer wave number $n$ corresponding to the 
quantized momentum along the extra large dimension. Since this sector is 
effectively higher-dimensional, the massive levels, $n\ne 0$, form multiplets 
of extended $N\ge 2$ supersymmetry (spontaneously broken to $N=0$). 
On the  other hand, the massless modes form multiplets of $N=1$ supersymmetry 
only, due to the orbifold projection. Finally, 
consistency of the theory implies in general the existence of a twisted sector 
corresponding to strings with centre of mass stacked at the orbifold fixed
points. It follows that the resulting states, which typically are chiral, 
live on four-dimensional boundaries of the higher-dimensional theory. Thus, 
they are not accompanied by KK excitations and form multiplets of $N=1$ 
(unbroken) supersymmetry.

It is now clear that the requirement for avoiding the large coupling problem,
already before supersymmetry breaking, is to impose the vanishing of the
contributions to the beta-functions of the massive untwisted tower of states,
level by level. This is sufficient to be done at the one-loop level, since the
effective $N\ge 2$ supersymmetry guarantees the absence of higher-loop
corrections \cite{ant}. The simplest solution is when KK modes come into 
multiplets of $N=4$ supersymmetry \cite{a,b} (possibly spontaneously broken 
to $N\le 2$ \cite{kkpr}). Since $N=4$ multiplets contain spin-1 fields, quark 
and lepton multiplets should come from the twisted sector due to 
phenomenological constraints, such as to avoid fast proton decay. The Higgs 
multiplet may originate from either sector, depending on model building 
\cite{amq}. When it is untwisted, the Standard Model gauge group is enlarged 
in the higher-dimensional theory \cite{ab}, while, when it is 
twisted, the KK excitations have just the quantum numbers of the $SU(3)\times
SU(2)\times U(1)$ gauge group.\footnote{In general, additional constraints may
have to be imposed to avoid potential growing of the Yukawa couplings.}

Since the presence of the infinite massive tower of $N=4$ KK multiplets does
not contribute to the beta-functions, the gauge couplings continue the
logarithmic evolution, as determined by the usual $N=1$ beta-functions. 
Inclusion of supersymmetry breaking gives negligible threshold effects,
exponentially suppressed in the large radius limit \cite{a}. As a result, the 
value of the unification scale inferred from low energy data does not change. 
A possible worry may arise from string non-perturbative dynamics. In fact,
although in these heterotic string models all four-dimensional couplings 
remain perturbative well above the decompactification scale, the 
ten-dimensional string coupling is strong, because of the large volume of the 
internal compactification space, which might be a source of stringy 
non-perturbative effects. These effects can be studied using duality and give 
rise to energy thresholds of the weakly coupled dual theory, which is type I
string theory or M-theory, above which the effective field theory description 
breaks down \cite{ckm}. In the case of one large dimension, the thresholds of
the dual theories turn out to be of the order of the unification scale, 
$\sim 10^{16}$ GeV, and are thus harmless for low energy physics and gauge 
coupling unification \cite{aq}.

The simplest way of breaking supersymmetry spontaneously by compactification 
is the Scherk-Schwarz mechanism \cite{ss,sss}. Higher-dimensional fields, 
instead of remaining periodic under a $2\pi$-rotation 
around the compact dimension, are allowed to pick up a (discrete) phase that 
can be absorbed by an R-symmetry transformation. This change of the boundary 
condition amounts to shifting the momentum KK number $n$ by the (discrete) 
R-charge, which splits the supersymmetry multiplets. Note that only the 
untwisted sector of the theory feels this breaking, while twisted states do 
not carry internal momentum and remain untouched to lowest order. 
Communication of supersymmetry breaking arises through gauge interactions and 
can be studied in the context of the effective field theory \cite{amq}. Since 
all mass splittings are of the order of the compactification scale, the scale 
of supersymmetry breaking is set by the same scale $R^{-1}$, which is at the 
TeV range. 

This situation is very similar to that of spontaneously broken global 
supersymmetry. It is different from the situation where supersymmetry 
breaking originates from a new strongly interactive sector; the 
breaking scale is thus set by the corresponding QCD-like dynamical scale 
$\Lambda_s$. In the case of gauge-mediated models, due to the gauge coupling 
suppression, $\Lambda_s$ is at least of the order of 10--100 TeV \cite{gm}, 
while for  gravity-mediated models $\Lambda_s$ is at an intermediate scale 
of order $10^{13}$ GeV. We will see below that this difference has very 
important consequences.

A natural question arising in models with weak-scale compactification is to
know which mechanism fixes the size of the extra dimension at the TeV$^{-1}$ 
scale. One possibility would be some strong supersymmetric dynamics at a higher
scale. In fact, as mentioned in the introduction, the value of $R$ corresponds
to a classical flat direction, which parametrizes the vacuum expectation value
(VEV) of a modulus scalar field originating from the diagonal component of the
metric along the extra dimension. The perturbative flatness could be lifted by
non-perturbative effects. Such effects can either have string origin, or can be
due to field theoretical strong gauge dynamics. However, the latter seems
impossible since all low energy couplings are independent from $R$ before
supersymmetry breaking, by construction. Furthermore, the former can be studied
by going to the weakly coupled dual theory, as we mentioned before. 
It follows that as long as the radius modulus remains flat in the dual theory, 
its VEV cannot be fixed by supersymmetric non-perturbative dynamics.

Alternatively, $R$ is fixed only after taking into account the 
supersymmetry-breaking effects, which can drive electroweak symmetry breaking 
radiatively \cite{rb}. The compactification scale could then be set by a new 
dynamical scale defined as the energy where the mass-squared of the Higgs 
becomes negative and leads to the breaking of the electroweak gauge symmetry. 
This mechanism, proposed in no-scale models \cite{nos}, could be realized 
provided there are no quadratic divergences in the vacuum energy. As we will 
see below, this condition is fulfilled in our case and a precise study of this 
possibility deserves further investigation.

The main prediction of weak scale compactification models for high energy
experiments is the existence of massive towers of KK modes with the quantum
numbers of the Standard Model gauge bosons. These modes are unstable as they
can decay into quarks and leptons within a short lifetime of the order of
$10^{-26}$ s \cite{ab}. Their clear experimental signature is the direct 
production, through for instance Drell-Yan processes in $pp$ and $p{\bar p}$ 
collisions \cite{abq,km}. 
At present energies, they lead only to small indirect
effects, such as effective dimension-six four-fermion operators. The resulting
limits come from bounds on compositeness and allow a compactification scale of
one dimension to be as low as a few hundred GeV \cite{ks,ab}.

\section{Analogy with finite temperature}

In order to be more explicit, without loss of generality for our conclusions, 
let us consider the simplest case of Scherk-Schwarz compactification of one 
large dimension using a $Z_2$ R-parity. It can be shown that such a generic 
and model-independent symmetry is the ordinary fermion number restricted to 
the untwisted sector of the theory \cite{a}. In this case all bosonic 
excitations are unaffected, while KK fermions split, since they become 
antiperiodic and obtain half-integer frequencies.
It follows that the mass spectrum of the tower of excitations which accompanies 
the massless states becomes:
\ba
\rm{vector\ multiplets}\quad &:&\qquad\qquad
M^2_B={n^2\over R^2}\qquad ;\qquad M^2_F={(n+{1\over 2})^2\over R^2}
\nonumber\\
\rm{twisted\ chiral\ multiplets}\quad &:&\qquad\qquad M^2_T=0\, .
\label{mass}
\ea

Note that the above breaking is identical to the one obtained at finite 
temperature by replacing time with a space coordinate and identifying the
temperature $T\equiv R^{-1}$ in the five-dimensional theory. Using this
analogy, we can understand some extremely soft ultraviolet properties of
supersymmetry breaking in these models, such as the absence of quadratic
divergences, or equivalently the vanishing of the supertrace
Str$M^2$ in the effective potential. This is consistent with the fact that all
soft breaking masses are in the TeV region. In fact, the vacuum energy density 
$E$ in these models behaves as \cite{it,a}:
\be
E\sim (n_F -n_B){1\over R^4} + {\cal O}(e^{-R^2})\, ,
\label{E}
\en
where $n_F$ and $n_B$ denote, respectively, the number of massless fermions and
bosons after supersymmetry breaking. This result is independent of the
requirement of $N=4$ supersymmetry among the massive KK excitations, and can be
understood as the analogue of the $T^4$ behaviour of the free energy at finite
temperature.

To illustrate this phenomenon, let us consider the contribution of a single
(five-dimen\-sional) multiplet to the vacuum energy, in the field theory limit
\cite{aq2}:
\ba
E &=& {1\over 2}{\rm Str}\int{d^4k\over (2\pi)^4}\ln (k^2+M^2)
=-{1\over 2}{\rm Str}\int{d^4k\over (2\pi)^4}\int_0^\infty{ds\over s}
e^{-s(k^2+M^2)}\nonumber\\
&=& -{1\over 32\pi^2}\int_0^\infty{ds\over s^3}\sum_n\left[
e^{-sn^2/R^2}-e^{-s(n+{1\over 2})^2/R^2}\right]\nonumber\\
&=& -{1\over 32\pi^2}\int_0^\infty{ds\over s^3}
\left({\pi R^2\over s}\right)^{1/2}
\sum_m\left[ 1-(-)^m\right] e^{-\pi^2 R^2/s}\nonumber\\
&=& -{93\, \zeta(5)\over 2^{10}\pi^6}{1\over R^4}\, ,
\label{E1}
\ea
where $\zeta(5)\simeq 1.037$. From the second line of
eq.~(\ref{E1}), it follows that every single multiplet in the sum gives a
quadratically divergent contribution which, in the proper time representation,
behaves as $ds/s^2$ in the ultraviolet limit $s=0$. The quartic divergence
$ds/s^3$ cancels among bosons and fermions of the same multiplet, as expected.
However, after summing over all modes the quadratic divergence disappears as
well, which is clear from the convergent integral in the third line of
eq.~(\ref{E1}), where we performed a Poisson resummation. In other words, the
contribution to the Str$M^2$ of the four-dimensional effective field theory
describing the $n=0$ mode is cancelled non-trivially by the infinite sum over
the massive KK excitations with $n\ne 0$.

By analogy with finite temperature, the ultraviolet softness of these theories
can also be understood by using an alternative description of Feynman 
propagators that makes manifest the Boltzman factors suppression at energies 
higher than the temperature. After summing over all KK excitations, the bosonic 
and the (square of) fermionic propagators yield:
\ba
\Delta_B &=& \sum_n{1\over k^2+n^2/R^2}={\pi R\over k}{\rm cth}(\pi Rk)
={\pi R\over k}\left\{ 1+{\cal O}(e^{-2\pi Rk})\right\}\, ,
\nonumber\\
\Delta_F &=& \sum_n{1\over k^2+(n+{1\over 2})^2/R^2}=
{\pi R\over k}{\rm th}(\pi Rk)
={\pi R\over k}\left\{ 1+{\cal O}(e^{-2\pi Rk})\right\}\, ,
\label{prop}
\ea
where the leading behaviour in the ultraviolet limit is dictated by the
five-dimensional supersymmetric field theory and the subleading corrections are
exponentially suppressed. It follows that the short-distance limit of the
difference of the two propagators is finite,
\be
\int{d^4k\over (2\pi)^4}(\Delta_B - \Delta_F)<\infty\, ,
\label{dif}
\en
which shows that no new divergences are introduced after supersymmetry
breaking. In particular, the vacuum energy behaves as $1/R^4\equiv T^4$, which 
is determined by naive dimensional analysis. 

At higher-loop level, an additional complication arises from the propagation
of matter twisted fields that have no KK modes, but acquire soft scalar
masses through radiative corrections. Then, the analogy with finite 
temperature holds only for graphs involving fields from the untwisted sector 
of the theory. Of course, graphs with only twisted fields give vanishing
contribution by supersymmetry. There remain the potentially dangerous diagrams 
with mixed propagation, which need more careful analysis. Consider for instance
the two-loop supergraph containing a loop of a (massless) twisted state 
exchanging an untwisted mode from the tower of KK excitations. This diagram
can be computed in two steps. 

As a first step, we consider the one-loop correction to the two-point 
function of the untwisted mode, due to the propagation
of the twisted fields. Obviously, this is a supersymmetric correction to the
wave function renormalization of the KK mode, which makes, in particular, the
coupling of all excited gauge bosons to run with the energy \cite{abq}. As a
second step, we compute the contribution to the vacuum energy of the
untwisted state --with one-loop corrected wave function-- and perform the sum
over all KK modes (\ref{mass}):
\be
E^{(2)}_{\rm mixed}={1\over 2}\gamma_Tg^2{\rm Str}V_M^2\int 
{d^4k\over (2\pi)^4}{k^2\ln k^2\over k^2+M^2}\, ,
\label{E2}
\en
where $\gamma_T$ is a numerical coefficient coming from the one-loop
integration over the twisted states, $g$ is the four-dimensional string
coupling, and $gV_M$ is the three-point vertex of the untwisted field with two
twisted states. For generic orbifold, $|V_M|=\delta^{M^2}$, with $\delta\le 1$
a model-dependent constant related to the order of the orbifold twist
\cite{ab}. In the large radius limit $M^2\to 0$ and for the leading 
contribution one can substitute $V_M=1$, since the
sum over KK momenta $n$ converges in eq.~(\ref{E2}).
Using now the mass formula (\ref{mass}) and introducing the proper time
representation (\ref{E1}), eq.~(\ref{E2}) becomes in the large $R$ limit:
\ba
E^{(2)}_{\rm mixed} &\sim &\partial_\varepsilon|_{\varepsilon=0}
{\rm Str}\int{d^4k\over (2\pi)^4}(k^2)^{1+\varepsilon}
\int_0^\infty ds e^{-s(k^2+M^2)}\nonumber\\
&\sim &\partial_\varepsilon|_{\varepsilon=0}
\int_0^\infty{ds\over s^{3-\varepsilon}}\sum_n
\left [e^{-sn^2/R^2}-e^{-s(n+{1\over 2})^2/R^2}\right]\nonumber\\
&\sim &\partial_\varepsilon|_{\varepsilon=0}
\int_0^\infty{ds\over s^{3-\varepsilon}}\left({\pi R^2\over s}\right)^{1/2}
\sum_m\left[ 1-(-)^m\right]e^{-\pi^2R^2/s}\nonumber\\
&\sim &{1\over R^4}\ln R\, .
\label{E21}
\ea

We will now show that the $\sim 1/R^4$ behaviour of the vacuum energy (up to
logarithms) can in fact be understood as a result of the (global) supersymmetry
algebra. In the globally supersymmetric limit, supersymmetry is spontaneously
broken through the boundary conditions, and the goldstino is identified with the
component of the five-dimensional gravitino along the extra dimension. In the
$N=1$ theory, there is only one zero mode that survives the orbifold
projection, and the non-linearity in its 
transformation, which measures the supersymmetry breaking, is \cite{ss}:
\begin{equation}
\delta \psi \sim {1\over\kappa_5 R}\epsilon\,
\sim\, R^{-3/2}\epsilon\, ,
\label{deltapsi}
\end{equation}
where $\kappa_5=R^{1/2}M_p^{-1}$ is the five-dimensional gravitational
coupling, and $\epsilon$ is the supersymmetry transformation parameter.
{}From the supersymmetry algebra then follows that the vacuum energy density 
of the theory must behave with the 3rd power of the gravitino mass, at least,
which implies $E\sim R^{-4}$ due to the evenness of the fermion number.

It should be stressed that the above behaviour of the vacuum energy holds in
spite of the fact that sparticle masses are hard. In models with gauge-mediated
supersymmetry breaking soft masses vanish with a power low above the messengers
mass. On the contrary, in our case the mass shifts
are hard, since they arise from a change of boundary conditions, and their
behaviour is determined in the supersymmetric theory through the wave-function
renormalizations. In the particular class of models we consider, all couplings
and sparticle
masses run logarithmically all the way up to the unification scale. For
instance, gaugino masses run through the evolution of gauge couplings, which is
unaffected by the presence of the infinite tower of massive excitations
due to the effective $N=4$ supersymmetry.

\section{Moduli masses and millimetre range forces}

We now study the question of moduli masses and their couplings to matter in
theories with weak-scale compactification. As we already mentioned, all these 
theories contain a universal scalar modulus whose VEV determines the large 
radius $R$. Furthermore, as we argued in Section 2, its flatness is lifted only
after supersymmetry breaking through a potential generated by radiative
corrections. To lowest order, its mass can be read off from eq.~(\ref{E1}), 
after taking into account that the field that has canonical kinetic terms is 
$\phi=\ln R$, in units of the reduced Planck mass $M_p$, as follows by direct 
dimensional reduction of the five-dimensional Einstein action:
\be
m_\phi = \left({93\, \zeta(5)\over 32\pi^6}N\right)^{1/2}{R^{-2}\over M_p}\, .
\label{mphi}
\en
Here, $N$ is the number of massless untwisted multiplets (before supersymmetry
breaking). Counting the Standard Model gauginos together with the higgsinos and
the gravitino, one finds $N\simgt 16$, which yields:
\be
m_\phi\simgt 0.22\ {R^{-2}\over M_p}\, .
\label{mphiv}
\en

It is a general property of moduli that they are gravitationally coupled
with their interactions to matter arising through the dependence of the low
energy couplings. Since in the absence of supersymmetry breaking all 
dimensionless couplings are independent of the radius modulus, its coupling to
matter is extremely suppressed. In the presence of supersymmetry
breaking, it acquires direct couplings through the soft masses, which again 
lead to suppressed interactions. It turns out that the dominant coupling comes
through the logarithmic running of gauge couplings after supersymmetry breaking.
This dependence arises as follows. 

Consider the nucleon mass term in the low energy effective Lagrangian:
\be
{\cal L}_N=m_N(\phi){\bar N}N\, ,
\label{N}
\en
where the nucleon mass $m_N(\phi)$, which depends on the  
canonically normalized radius modulus $\phi=\ln R$, is proportional to  
the QCD scale $\Lambda_{\rm QCD}$. 
Since the sparticle masses are all proportional to  
$1/R$, the sparticle threshold  $\Lambda_{\rm SUSY}$ is proportional to  
$1/R$. The coupling of the radius modulus to nucleons is therefore  
given by the derivative of the nucleon mass relative to the modulus  
$\phi$:
\be
{\partial m_N\over\partial\phi}=
{\partial\Lambda_{\rm QCD}\over\partial\ln\Lambda_{\rm SUSY}}\, .
\label{Np}
\en
Since the graviton coupling to the nucleon is $m_N/M_p$, the coupling $\alpha$
of the modulus relative to gravity is
\be
\alpha_\phi = {\partial\ln\Lambda_{\rm QCD}\over\partial\ln\Lambda_{\rm SUSY}}
= 1-{b_{SS}\over b_{NS}}={4\over 7}\, ,
\label{coupling}
\en
where $b_{SS}$ is the supersymmetric beta-function, which includes  
all the sparticles, and $b_{NS}$ is the non-supersymmetric beta-function, 
which includes just the ordinary particles. In this formula  
we assumed for simplicity that all sparticles have the same mass $\sim  
\Lambda_{\rm SUSY}$.
Thus we see that the force between two pieces of matter mediated by  
the radius modulus is $(4/7)^2 \simeq 1/3$ times the force of gravity.  

In principle there can be other light moduli which couple with larger  
strengths. For example the dilaton, whose VEV determines the (logarithm of the)
string coupling constant, if it does not acquire large mass from some dynamical
supersymmetric mechanism, leads to the strongest effect.
Its coupling is \cite{tv}:
\be
\alpha_{\rm dilaton}=\ln{M_p\over\Lambda_{\rm QCD}}\sim 44\, ,
\label{alphadil}
\en
which corresponds to a strength $\sim 2000$ times bigger than gravity.  
Therefore, a generic prediction of these theories is the existence of  
moduli with Compton wavelengths in the millimetre range and with  
couplings ranging from $\sim$ gravitational to several times larger. 
The important point we want to emphasize here is  
that at least one of these, namely the radius modulus, appears to be a  
necessary consequence of the weak-scale compactification theories.

\begin{figure*}
\psfig{figure=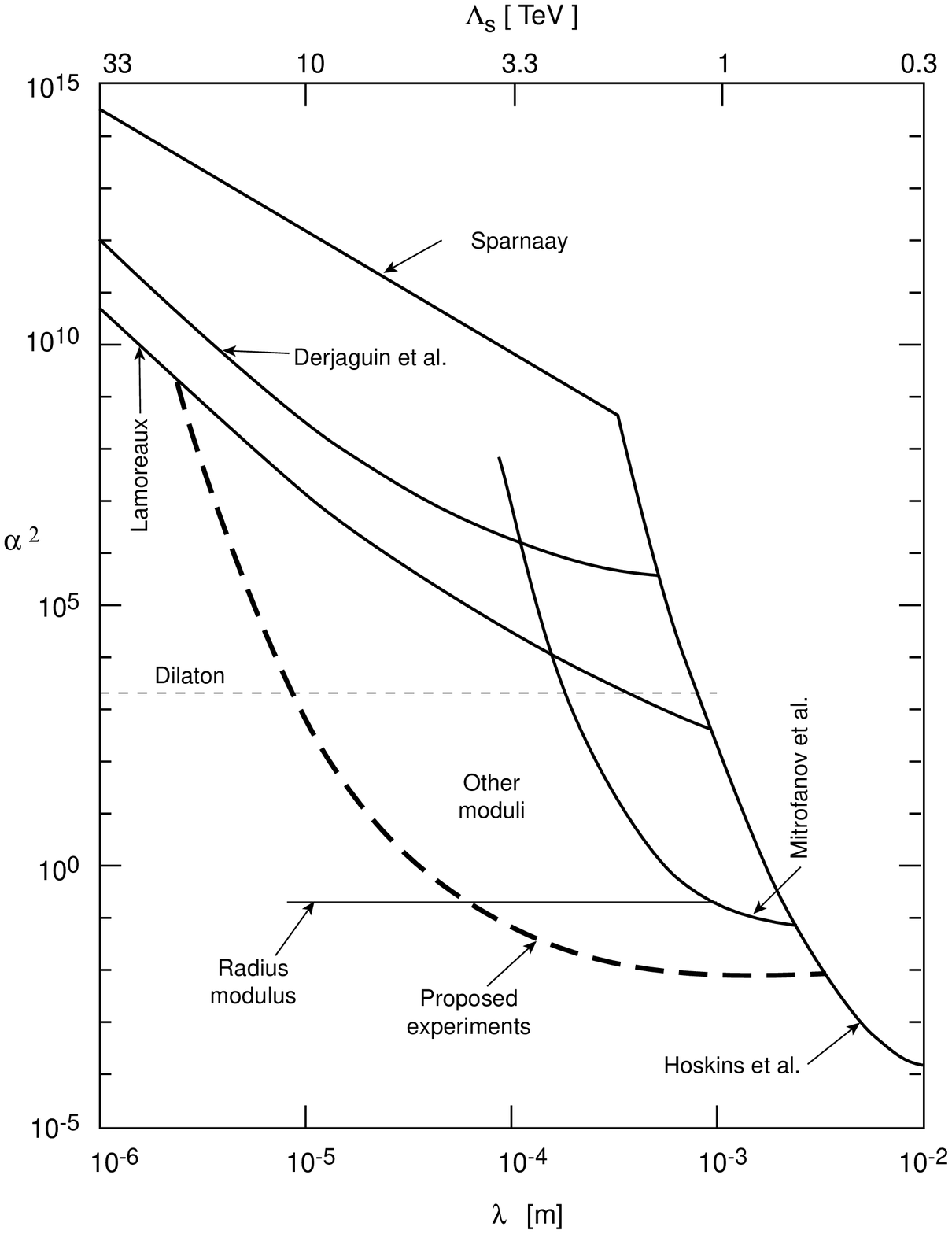,height=20cm,width=15cm}
Strength of the modulus force relative to gravity ($\alpha^2$) versus 
the Compton wavelength ($\lambda$) of the modulus. The upper scale shows the
corresponding value of the large radius in TeV.
\end{figure*}

In the figure we depict our theoretical predictions together with information 
from present and upcoming experiments. The vertical axis is the strength, 
$\alpha^2$, of the force relative to gravity; the horizontal axis is the 
Compton wavelength of the exchanged particle; the upper scale shows the
corresponding value of the large radius in TeV.
The solid lines indicate the present limits from 
the experiments indicated \cite{oldexperiments}. The excluded regions lie 
above these solid lines. Our theoretical prediction for the radius modulus is 
the thin horizontal line at 1/3 times gravity. The dilaton-mediated force is 
the thin dashed horizontal line at $\sim$ 2000 times gravity. The proposed 
experiments are sensitive to dilatons of Compton wavelengths as small as 10 
$\mu$m. Other moduli may mediate forces between the above two extremes.

The most important part of the figure is the dashed thick line; it is the 
expected sensitivity of two heroic proposed experiments by Price et al. and 
by Kapitulnik et al. \cite{price,kapitulnik}.
They  use cantilever technology or  micro-electromechanical systems utilizing 
microsensors based on silicon micromachining. On the time-scale of a year they 
will improve the present limits by almost 6 orders of magnitude and --at the 
very least-- they will, for the first time, measure gravity to a precision of 
1\% at distances of $\sim$ 100 $\mu$m.

As we see from the figure, if the ideas presented in this paper are  
correct, these experiments have a good chance of finding the force  
mediated by the radius modulus. In fact, if the scale of  
supersymmetry breaking is at its most likely range of a TeV,
the force is just below the sensitivity of the Mitrofanov et al.  
experiment.  Actually, scales smaller than a TeV are already excluded  
from this experiment. If the radius modulus exists, then it becomes  
inaccessible to the proposed experiments only if the scale of  
supersymmetry breaking is higher than 5 TeV.
This scale is uncomfortably large  for the hierarchy problem: it  
requires fine tunings in excess of one part in $\sim 100$. It is also  
too large for the LHC to have a shot at finding supersymmetric  
particles.  It is therefore fair to conclude that if the proposed  
experiments do not find the force mediated by the radius modulus, they  
will shed strong doubts on the idea of weak-scale compactification.

\section{What is special about weak-scale compactification theories?}

There are key differences between weak-scale compactification and  
gauge- or gravity-mediated theories. One is that weak-scale  
compactification leads to {\it direct} supersymmetry breaking without  
the need for an intervening sector that feels primordial  
supersymmetry breaking. This is only possible because the underlying  
theory is higher-dimensional and therefore sidesteps the four-dimensional 
supertrace theorems which make direct supersymmetry breaking  
phenomenologically impossible in four dimensions  
\cite{fgp,dg}. It has the crucial advantage that  
the {\it largest} supersymmetry-breaking splittings in the full  
theory are at the weak scale. Consequently the Compton wavelength of  
the moduli, as given by eq.~(\ref{mmod}), is large.

A related unique theoretical feature of these theories is the  
simultaneous occurrence of {\it hard} gluino and squark masses  
$\Lambda_{\rm SUSY}\sim$ TeV together with vacuum energy that is  
$\Lambda_{\rm SUSY}^4\sim ({\rm TeV})^4$, i.e. the vacuum energy is at most  
logarithmically sensitive to the ultraviolet cutoff. This opens the  
possibility that the VEVs of some moduli may be predominantly  
determined by low energy (weak scale) physics. This suggests that the 
weak-scale compactification theories provide a natural home for the 
no-scale models. In fact the no-scale mechanism  
for fixing the scale of supersymmetry breaking, as the place where the Higgs  
mass-squared turns negative, seems to be the most economical way of  
dynamically determining the compactification radius. The guaranteed  
absence of strong ultraviolet sensitivity makes this possible.

Another application of this is the mechanism of dynamical alignment  
\cite{dimopoulosgiudiceunpublished}. This is the idea that the  
orientation of the soft terms in flavour space depends on moduli whose  
VEVs are not determined by ultraviolet physics. This implies that their VEVs,  
for energetic reasons, will line-up with the fermion masses for the  
same reason spins align with a magnetic field.  In fact, such a  
mechanism may be needed in these theories since the sfermion masses  
are hard and, consequently, can be distorted by Planckian flavour  
physics \cite{HKR}.

Another feature of weak-scale compactification is the  
existence of the large radius; this makes it extremely likely that the  
modulus corresponding to the large radius is massless until supersymmetry  
breaking takes over at the weak scale. Therefore, the existence of at  
least one light modulus seems unavoidable. In other words, if a force of
$\sim 1/3$ times gravity in the millimetre range is {\it not}  
measured, it will cast  serious doubt on the idea of weak-scale  
compactification.
 
Our most important conclusions are well summarized in the figure.  
There we see the future possibilities for the detection of sub-centimetre  
gravitational strength forces. These searches are important for two  
reasons. First they are, for the first time, going to detect gravity  
in a totally new range of distances. This is a sure thing, but very  
important nevertheless. Secondly, they may discover a new force of  
Nature which, according to the theories discussed here, will give us a  
first glimpse into extra dimensions of spacetime.
\vskip 0.5cm

\noindent{\bf Acknowledgements}

We would like to acknowledge very useful discussions and  
correspondence with  Gian Giudice, Joshua Long, John Price, Riccardo  
Rattazzi and, especially, Aharon Kapitulnik.


\end{document}